\documentclass[reprint,amsmath,amssymb,pra,superscriptaddress,floatfix,showkeys]{revtex4-1}
\usepackage{graphicx}
\usepackage[latin1]{inputenc}
\usepackage{amsmath,amsfonts,amssymb,url,hyperref}
\usepackage[american]{babel}
\hypersetup{colorlinks=true,breaklinks=true,urlcolor=blue,citecolor=blue,linkcolor=blue,pdfstartview=FitH,pdfpagemode=UseNone}

\usepackage[T1]{fontenc}
\usepackage{txfonts}

\begin{document}
\title{Three-dimensional splitting dynamics of giant vortices in Bose--Einstein condensates}
\author{Jukka R\"abin\"a}
\email{jukka.rabina@jyu.fi}
\affiliation{University of Jyvaskyla, Faculty of Information Technology, P.O. Box 35, FI-40014 University of Jyvaskyla, Finland}

\author{Pekko Kuopanportti}
\affiliation{Department of Physics, University of Helsinki, P.O. Box 43, FI-00014 Helsinki, Finland}

\author{Markus Kivioja}
\affiliation{University of Jyvaskyla, Faculty of Information Technology, P.O. Box 35, FI-40014 University of Jyvaskyla, Finland}

\author{Mikko M\"ott\"onen}
\affiliation{QCD Labs, QTF Centre of Excellence, Department of Applied Physics, Aalto University, P.O. Box 13500,
FI-00076 AALTO, Finland.}

\author{Tuomo Rossi}
\affiliation{University of Jyvaskyla, Faculty of Information Technology, P.O. Box 35, FI-40014 University of Jyvaskyla, Finland}
 
\date{\today}

\begin{abstract}
We study the splitting dynamics of giant vortices in dilute Bose--Einstein condensates by numerically integrating the three-dimensional Gross--Pitaevskii equation in time. By taking advantage of tetrahedral tiling in the spatial discretization, we decrease the error and increase the reliability of the numerical method. An extensive survey of vortex splitting symmetries is presented for different aspect ratios of the harmonic trapping potential. The symmetries of the splitting patterns observed in the simulated dynamics are found to be in good agreement with predictions obtained by solving the dominant dynamical instabilities from the corresponding Bogoliubov equations. Furthermore, we observe intertwining of the split vortices in prolate condensates and a split-and-revival phenomenon in a spherical condensate.
\end{abstract}


\maketitle

\section{\label{sec:intro}Introduction}

Quantized vortices are archetypal topological objects that play important roles in various branches of physics, ranging from superconductors~\cite{Par1969.book.superconductivity} and helium superfluids~\cite{Don1991.book.Vortices} to cosmology~\cite{Vil1994.book.cosmic_strings} and optics~\cite{And1975.Nat256.25}. Quantized vortices exist in matter fields described by a smooth complex-valued scalar field. The essential idea is that, while the complex field itself is single valued, its phase is defined only modulo $2\pi$. Hence, the contour integral of the phase around a closed loop need not vanish, but may in fact be any integer multiple $\kappa$ of $2\pi$. A nonzero $\kappa$ implies the presence of a quantized vortex within the loop and is referred to as the winding number of the vortex.

Bose--Einstein condensates (BECs) of atomic gases are dilute superfluids, which can be described by tractable theories~\cite{Dal1999.RMP71.463,Fet2009.RMP81.647} and are highly controllable in experiments~\cite{And2010.JLTP161.574}. Thus, they are excellent physical systems for studying quantized vortices. The BEC community has devoted a lot of attention to multiquantum vortices, for which $\lvert \kappa \rvert \geq 2$, and giant vortices, for which $\lvert \kappa \rvert >> 1$. Methods used to create them in gaseous BECs have so far included topological phase engineering~\cite{Lea2002.PRL89.190403,Shi2004.PRL93.160406,Iso2007.PRL99.200403,Kuw2010.JPSJ79.034004,Shi2011.JPhysB44.075302}, coherent transfer of angular momentum from photons to the atoms~\cite{And2006.PRL97.170406}, and removal of atoms from a lattice of single-quantum vortices by a tightly focused laser beam~\cite{Eng2003.PRL90.170405,Sim2005.PRL94.080404}. Given that the kinetic energy of a vortex is proportional to $\kappa^2$, a multiquantum vortex typically has a higher energy than a cluster of $\lvert \kappa \rvert$ separated singly quantized vortices. This makes multiquantum vortices prone to split into singly quantized vortices. The associated instabilities and dynamics have been studied both theoretically~\cite{Pu1999.PRA59.1533,Sim2002.PRA65.033614, Mot2003.PRA68.023611,Kaw2004.PRA70.043610, Huh2006.PRL97.110406,Mat2006.PRL97.180409,Gaw2006.JPhysB39.L225,Lun2006.PRA74.063620,Kar2009.JPhysB42.095301,Kuo2010.PRA81.023603,Kuo2010.PRA81.033627,Kuo2010.JLTP161.561,Cid2017.PRA96.023617} and experimentally~\cite{Shi2004.PRL93.160406,Iso2007.PRL99.200403,Kuw2010.JPSJ79.034004,Shi2011.JPhysB44.075302}. Recent studies have also addressed utilizing vortex splitting as a means to generate quantum turbulence with controllable net circulation~\cite{Abr1995.PRB52.7018,Ara1996.PRB53.75,Cid2016.PRA93.033651}. Besides being interesting due to their dynamics, multiquantum vortices could also be used to implement a ballistic quantum switch~\cite{Mel2002.Nat415.60} or realize bosonic quantum Hall states~\cite{Ron2011.SciRep1.43}. 

Previous theoretical studies of vortex splitting have been limited to relatively small winding numbers $\lvert \kappa \rvert \leq 5$~\cite{Pu1999.PRA59.1533,Sim2002.PRA65.033614,Mot2003.PRA68.023611,Kaw2004.PRA70.043610,Huh2006.PRL97.110406,Mat2006.PRL97.180409,Gaw2006.JPhysB39.L225,Lun2006.PRA74.063620,Kar2009.JPhysB42.095301,Cid2017.PRA96.023617} or to quasi-two-dimensional models pertaining to highly oblate BECs~\cite{Kuo2010.PRA81.023603,Kuo2010.PRA81.033627,Kuo2010.JLTP161.561,Cid2016.PRA93.033651}. In Ref.~\cite{Li2012.PRA86.023628}, vortex splitting was studied in three dimensions up to $\kappa=45$, but only for small BECs in isotropic harmonic traps. Splitting patterns exhibiting up to tenfold rotational symmetry were observed in the numerical simulations. In this work, we carry out a more comprehensive investigation of giant-vortex splitting in three-dimensional BECs. Considering all three different types of cylindrically symmetric harmonic traps (oblate, spherical, and prolate) and a wide range of repulsive interaction strengths, we simulate the temporal evolution of axisymmetric giant vortex states subjected to small random perturbations. In general, we find good agreement between the splitting patterns observed in the evolution and those predicted by linear stability analysis. Vortex splitting in prolate BECs is found to result in branched intertwining of the vortices, and spherical BECs are observed to exhibit a split-and-revival effect.

Importantly, we also find that the splitting patterns appearing in the simulated time evolution can be prone to numerical artifacts stemming from the symmetry of the underlying spatial grid. As a result, particular care should be taken when discretizing the time-dependent Gross--Pitaevskii equation (GPE) for the condensate. Specifically, the Cartesian grids used in the previous investigations tend to favor the fourfold splitting pattern, which may explain why, in Ref.~\cite{Kuo2010.PRA81.033627}, the higher-symmetry splitting patterns predicted by the linear stability analysis were not observed to arise from random perturbations. We solve this problem by basing our time integration scheme on discrete exterior calculus~\cite{2003hirani,2008desbrun,2015stern} with tetrahedral tiling.

The remainder of this article is organized as follows: In Sec.~\ref{sec:method}, we present the time-dependent GPE, derive the Bogoliubov equations used for the linear stability analysis, and outline our numerical integration method. Section~\ref{sec:results} begins with an analysis of the integration method and presents our numerical results. Finally, we conclude the paper in Sec.~\ref{sec:discussion}.

\section{\label{sec:method}Theory and Method}

\subsection{Mean-field model}

The complex-valued order parameter $\Psi$ of a dilute BEC at low temperatures satisfies the GPE
\begin{align*}
i \hbar \partial_t \Psi(\mathbf{r}, t) = \left[ -\tfrac {\hbar^2} {2 m} \nabla^2 + V(\mathbf{r}) + g |\Psi(\mathbf{r}, t)|^2 \right] \Psi(\mathbf{r}, t),
\end{align*}
where $i$ is the imaginary unit, $\hbar$ is the reduced Planck constant, $m$ is the atom mass, and $g$ is the effective interaction strength. The order parameter is normalized such that $\int |\Psi(\mathbf{r}, t)|^2 \mathrm{d}^3 r = N$ is the number of condensed atoms. We employ a cylindrically symmetric harmonic trapping potential $V(\mathbf{r}) =m \bigl( \omega_r^2 r^2 + \omega_z^2 z^2 \bigr)/2$, where $\omega_r$ and $\omega_z$ are the radial and axial trapping frequencies, respectively. 

To have generally applicable results, we employ dimensionless units and measure position in the units of the radial harmonic oscillator length $a_r = \sqrt{\hbar / m \omega_r}$, time in units of $1 / \omega_r$, the order parameter in units of $\sqrt{N / a_r^3}$, and the effective interaction strength in units of ${a_r^3 \hbar \omega_r} / N$. Thus, the conversion into the dimensionless units (denoted by a bar) is given by
\begin{align*}
\bar{\mathbf{r}} = \frac {\mathbf{r}} {a_r}, \quad \bar{t} = t \omega_r, \quad \bar{\Psi}(\bar{\mathbf{r}}, \bar{t}) = \Psi(\mathbf{r}, t) \sqrt{\frac {a_r^3} N}, \quad \bar{g} = \frac {g N} {a_r^3 \hbar \omega_r}.
\end{align*}
Consequently, the dimensionless order parameter is normalized as $\int |\bar{\Psi}(\bar{\mathbf{r}}, \bar{t})|^2 \mathrm{d}^3 \bar{\mathbf{r}} = 1$, and it satisfies the dimensionless GPE
\begin{align}
i \partial_{\bar{t}} \bar{\Psi}(\bar{\mathbf{r}}, \bar{t}) = \left[ -\tfrac 1 2 \bar{\nabla}^2 + \bar{V}(\bar{\mathbf{r}}) + \bar{g} |\bar{\Psi}(\bar{\mathbf{r}}, \bar{t})|^2 \right] \bar{\Psi}(\bar{\mathbf{r}}, \bar{t}). \label{equ_normalizedGPE}
\end{align}
The dimensionless potential is given by $\bar{V}(\bar{\mathbf{r}}) = \bigl( \bar{r}^2 + \lambda^2 \bar{z}^2 \bigr) / 2$, where $\lambda = \omega_z / \omega_r$ is referred to as the aspect ratio. In cylindrical coordinates, the Laplacian is given by $\bar{\nabla}^2 = \partial_{\bar{r}}^2 + \bar{r}^{-1} \partial_{\bar{r}} + \bar{r}^{-2} \partial_{\phi}^2 + \partial_{\bar{z}}^2$.

Equation~\eqref{equ_normalizedGPE} has stationary vortex solutions $\bar{\Psi}_{\lambda, \bar{g}, \kappa}$, which depend on $\lambda$, $\bar{g}$, and the integer winding number $\kappa$. These stationary states can be written as
\begin{align}
\bar{\Psi}_{\lambda, \bar{g}, \kappa}(\bar{\mathbf{r}},\bar{t}) &= f(\bar{r},\bar{z}) e^{i \kappa \phi - i \mu \bar{t}}, \label{equ_stationaryGPE}
\end{align}
where $f$ is a real-valued function and $\mu$ is the chemical potential. The stationary vortex states satisfy the time-independent equation
\begin{align*}
\left[ \tfrac{1}{2} \left( \tfrac {\kappa^2} {\bar{r}^2} - \partial_{\bar{r}}^2 - \tfrac 1 {\bar{r}} \partial_{\bar{r}} - \partial_{\bar{z}}^2 \right) + \bar{V} + \bar{g} f^2 \right] f = \mu f,
\end{align*}
which can be solved using a relaxation method~\cite{1970ortega}.

\subsection{Bogoliubov equations and stability}

To study the local stability properties of a given stationary vortex solution $\bar{\Psi}_{\lambda, \bar{g}, \kappa}$, we decompose the order parameter as
\begin{align}
\bar{\Psi}(\bar{\mathbf{r}}, \bar{t}) = \left[ f(\bar{r},\bar{z})+\chi(\bar{\mathbf{r}},\bar{t}) \right] e^{i\kappa\phi -i \mu \bar{t}},
\label{eq:decomposition}
\end{align}
where $\chi$ is a function describing a small perturbation such that $\int |\chi(\bar{\mathbf{r}},\bar{t})|^2 \mathrm{d}^3 \bar{r} \ll  1$. 
By substituting Eq.~\eqref{eq:decomposition} into Eq.~\eqref{equ_normalizedGPE}, neglecting the second- and third-order terms in $\chi$, and seeking oscillatory solutions of the form
\begin{align}
\chi(\bar{\mathbf{r}}, \bar{t}) = \sum_{q\in\mathbb{N}}\sum_{l\in\mathbb{Z}} \left[ u_{q,l}(\bar{r},\bar{z}) e^{ i l \phi-i\omega_{q,l} \bar{t}} + v_{q,l}^\ast(\bar{r},\bar{z}) e^{i \omega_{q,l}^\ast \bar{t} - i l \phi} \right], \label{eq:oscillation}
\end{align}
we obtain the Bogoliubov equations
\begin{align}
\begin{pmatrix} \mathcal{M}_{l\,} & \bar{g} f^2 \\ -\bar{g} f^2 & -\mathcal{M}_{-l\,} \end{pmatrix} \begin{pmatrix} u_{q,l} \\ v_{q,l} \end{pmatrix}
 = \omega_{q,l} \begin{pmatrix} u_{q,l} \\ v_{q,l} \end{pmatrix}, \label{eq:axisym2cBogoeq}
\end{align}
where the linear differential operator is defined as
\begin{align*}
\mathcal{M}_{l} = \tfrac 1 2 \left[ \tfrac{\left(\kappa+l\right)^2}{\bar{r}^2} - \partial^2_{\bar{r}} - \tfrac 1 {\bar{r}} \partial_{\bar{r}} - \partial^2_{\bar{z}} \right] + \bar{V} + 2 \bar{g} f^2-\mu.
\end{align*}
The integer $l$ specifies the angular momentum of the excitation with respect to the condensate, and $q\in\mathbb{N}$ is an index for the different eigenmodes with a given $l$. 

Equation~\eqref{eq:axisym2cBogoeq} can be used to determine the stability characteristics of the stationary vortex state in question. If the excitation spectrum $\{\omega_{q,l}\}$ contains at least one eigenfrequency with a positive imaginary part $\mathrm{Im}(\omega_{q,l})>0$, the state is \emph{dynamically unstable}; otherwise, the state is dynamically stable. If the spectrum contains an excitation for which $\mathrm{Re}(\omega_{q,l}) < 0$ and $\iint \bigl(|u_{q,l}|^2-|v_{q,l}|^2 \bigr) \bar{r}\,\mathrm{d}\bar{r} \,\mathrm{d}\bar{z} \ge 0$, the state is \emph{energetically unstable}; if no such excitations exist, the stationary state is (locally) energetically stable. We emphasize that energetic stability is a stronger condition than dynamical stability, since the former implies the latter.

As can be observed from Eq.~(\ref{eq:oscillation}), the occupations of excitation modes with $\mathrm{Im}(\omega_{q,l})>0$ are predicted to increase exponentially over time, and, consequently, small perturbations of a dynamically unstable stationary state typically lead to large changes in its structure. For dynamically unstable multiquantum vortices, in particular, the complex-frequency modes usually induce instability against \emph{splitting} of the multiply quantized vortex into singly quantized ones. In fact, the quantity $\max_{l}\max_q[\mathrm{Im}(\omega_{q,l})]/2\pi$ and the maximizing winding number $l$ can be used to predict, respectively, the inverse lifetime of a vortex and the symmetry of its typical splitting pattern~\cite{Mot2003.PRA68.023611}. Note, however, that the dynamically unstable modes quickly drive the system beyond the linear regime of the Bogoliubov analysis. As a result, the long-time dynamics of dynamically unstable states must be described with the time-dependent GPE, Eq.~\eqref{equ_normalizedGPE}, instead.

\subsection{Time integration}

Finite-difference methods have become popular for solving the time-dependent GPE because of their simplicity~\cite{Cer2000.PRE62.1382, 2015moxley, 2016young}. Alternative spectral methods~\cite{2003bao, 2006bao, 2013antoine, Dio2003.PRE67.046706} are also widely used. Typically, these methods rely on Cartesian spatial discretization, even though there are strong reasons to prefer simplicial grids~\cite{2004bossavit,2015rabina-siam}.

This work, on the contrary, utilizes a time integration method based on discrete exterior calculus (DEC)~\cite{2003hirani, 2008desbrun, 2015stern}, which naturally segregates the differentiable and metric structures~\cite{1999bossavit, 2000bossavit}. This approach can be regarded as a generalized finite-difference technique that closely resembles the finite integration technique~\cite{2003weiland} or the finite-difference time-domain method~\cite{1966yee, 1980taflove}. The DEC method is applicable to unstructured grids, while being stable and conserving the particle number.

The discretization is based on a pair of interlocked three-dimensional meshes: a primal (Delaunay) mesh and its dual (Voronoi) mesh. We assign each dual node with a floating point number to obtain a column vector $\psi^k$ that represents the discrete order parameter at a time instance $k {\Delta t} / 2$, where $k$ is an integer and $\Delta t$ is the length of the time step. With the notation of Ref.~\cite{2018rabina-m2an}, the discrete Laplacian is denoted as $\star_3 {\rm d}_2 \star_2^{-1} {\rm d}_2^T$, where $\star_p$ is a diagonal matrix called the \emph{discrete Hodge} and ${\rm d}_2$ is a sparse matrix called the \emph{discrete exterior derivative}. The time integration of Eq.~\eqref{equ_normalizedGPE} is carried out using the central-difference method
\begin{align*}
\psi^{k+1} &= \psi^{k-1} - i {\Delta t} \left( \star_3 {\rm d}_2 \star_2^{-1} {\rm d}_2^T / 2 + {\rm V}^k \right) \psi^k,
\end{align*}
where ${\rm V}^k$ is a diagonal matrix with elements ${\rm V}_{jj}^k = \bigl( \bar{r}_j^2 + \lambda^2 \bar{z}_j^2 \bigr) / 2 + \bar{g} \lvert \psi_j^k \rvert^2$. Here $\bar{r}_j$ and $\bar{z}_j$ denote the radial and axial coordinates of the $j$th dual node. The method is numerically stable if ${\Delta t} < M^{-1}$, where $M$ is the maximal diagonal element of the matrix $\star_3 {\rm d}_2 \star_2^{-1} {\rm d}_2^T / 2 + {\rm V}^k$.

\section{\label{sec:results}Results}

\subsection{\label{sec:results1}Evaluation of time integration} 

First, we test our numerical solver by numerically integrating a stationary vortex state forward in time and investigating its stability during the simulation. We consider the normalized GPE~\eqref{equ_normalizedGPE} with parameters $\lambda = 1$, $\bar{g} = 300$, and $\kappa = 10$. The time integrator is initialized at time instances $-{\Delta t} / 2$ and 0 by letting $\psi_j^k = \bar{\Psi}_{\lambda,\bar{g}, \kappa} (\bar{\mathbf{r}}_j, k {\Delta t})$, where $k = -1, 0$ and $\bar{\mathbf{r}}_j$ is the $j$th dual node position of the mesh. 

\begin{figure}[htb]
\includegraphics[width=0.9\columnwidth]{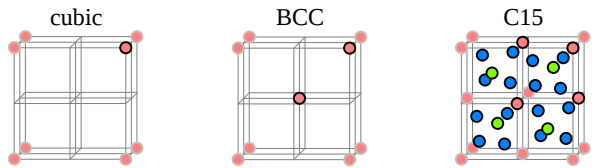}
\caption{Node positions for cubic, BCC, and C15 tilings.}
\label{fig_grids3d}
\end{figure}

Let us vary the spatial mesh and consider its effects on the solution. We employ three qualitatively different grids, which correspond to Delaunay meshes generated by the node positions illustrated in Fig.~\ref{fig_grids3d}. The simplest and most commonly used grid is the one with the cubic tiling. Its popularity is mainly based on its ease of implementation. Second, we employ body-centered cubic (BCC) tiling~\cite{1922sommerville, 1999conway}, which is preferred by certain numerical studies~\cite{2004bossavit, 2008vanderzee}. The third option is the C15 structure, which is one of the tetrahedrally close-packed tilings~\cite{1927friauf, 2011paufler, 1999sullivan, 2004eppstein}. The C15 structure has been found to be a high-quality grid for the solution of the Maxwell equations~\cite{2015rabina-siam, 2018rabina-m2an}. For each of these three grid types, we employ three discretization levels, where tasks are scaled to involve $10^9$, $10^{10}$, or $10^{11}$ floating point multiplications for integration over a unit time interval. 

\begin{figure}[htb]
\includegraphics[width=\columnwidth]{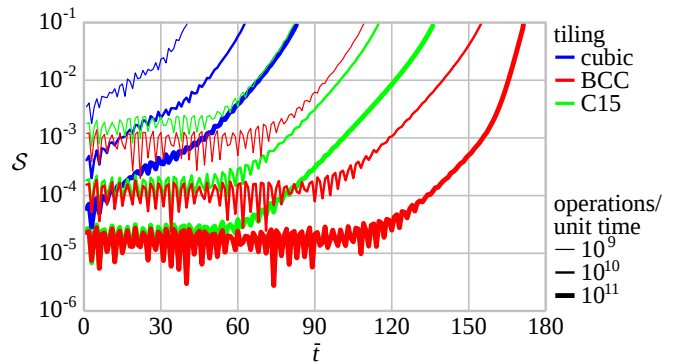}
\caption{Error $\mathcal{S}(\bar{t})$ induced by the numerical implementation of the GPE as a function of time for different tilings and discretization levels. The parameters for the stationary state are $\lambda = 1$, $\bar{g} = 300$, and $\kappa = 10$.}
\label{fig_stability}
\end{figure}

During the integration, we monitor the deviation $\mathcal{S}(\bar{t}) = 1 - \left| \int \bar{\Psi}_{\lambda,\bar{g}, \kappa}^*(\bar{\mathbf{r}}, 0) \bar{\Psi}(\bar{\mathbf{r}}, \bar{t}) \, {\rm d}^3 \bar{r} \right|$ from the stationary state and terminate the simulation when $\mathcal{S}(\bar{t})$ exceeds $0.1$. The duration before the termination is referred to as the \emph{time span of stability}. The evolution of $\mathcal{S}(\bar{t})$ is illustrated in Fig.~\ref{fig_stability}.

The time span of stability appears to be very sensitive to the grid type used. The BCC grid offers the longest time spans, since it is numerically the most isotropic of the three grids~\cite{2018rabina-m2an}. With the finest discretization level, BCC leads to threefold splitting, which is the most likely physical solution for the used parameter values (see Sec.~\ref{sec:results2}). In other cases, the fourfold symmetry of the cubic base grid steers the numerical solution into fourfold splitting. This demonstrates the importance of the tiling in obtaining correct physical results.

The BCC grid also offers the smallest early-stage errors before the actual vortex splitting occurs. The early-stage error seems to approximately obey the function $h^4$, where $h$ is the dual edge length. With the lowest discretization level ($10^9$ operations/unit time), the average dual edge lengths are $0.20$, $0.16$, and $0.17$ for the cubic, BCC, and C15 grids, respectively. The edge lengths of the finest ($10^{11}$) and second finest ($10^{10}$) discretization levels are about 0.38 and 0.61 times the above-mentioned edge lengths, respectively. 

Owing to these results, we choose to employ the BCC grid in the remaining numerical simulations presented in this work.

\begin{figure*}
\includegraphics[width=2\columnwidth]{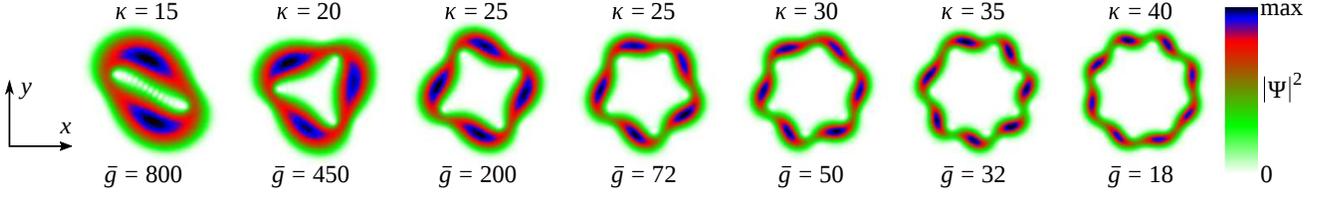}
\caption{Particle density $|\bar{\Psi}(\bar{t})|^2$ in an oblate BEC ($\lambda = 10$) integrated over $z$ at time $\min \{ \bar{t} \; | \; \mathcal{P}_{l_{\rm dom}}(\bar{t}) > 0.2 \}$ (see text for the definition of $l_{\rm dom}$). Typical $l_{\rm dom}$-fold splitting patterns appear for $l_{\rm dom} = 2,3,...,8$, respectively.}
\label{fig_typical}
\end{figure*}

\subsection{\label{sec:results2}Dominant splitting symmetries}

Even the smallest random perturbation to a dynamically unstable stationary vortex state triggers the splitting of the vortex. To find the most likely physical splitting symmetries, the stationary vortex states are perturbed slightly by adding low-amplitude random noise in the beginning of the computation. The discrete order parameter is initialized at instances $k = -1, 0$ by
\begin{align*}
\psi_j^k =  \left( 1 + \tfrac 1 {10} \rho_j \right) \bar{\Psi}_{\lambda,\bar{g}, \kappa} (\bar{\mathbf{r}}_j, k {\Delta t}),
\end{align*}
where $\rho_j$ is a random variable chosen uniformly from the unit disk in the complex plane.

The spatial discretization employs the BCC grid, whose dual edge lengths are $< 5\%$ of the effective wavelength 
\begin{align*}
\mathcal{L}_{\lambda,\bar{g}, \kappa} = \frac {2 \pi} {\sqrt{-\int \bar{\Psi}_{\lambda,\bar{g}, \kappa}^*(\bar{\mathbf{r}}, 0) \nabla^2 \bar{\Psi}_{\lambda,\bar{g}, \kappa}(\bar{\mathbf{r}}, 0) {\rm d}^3 \bar{r}}}.
\end{align*}
This corresponds to the second finest discretization level of Sec.~\ref{sec:results1}. The computational domain is a rectangle that contains all points $\bar{\mathbf{r}}$ for which $|\bar{\Psi}_{\lambda,\bar{g}, \kappa}(\bar{\mathbf{r}},0)|$ is greater than $10^{-5}$ times its maximum. Zero particle density is employed as the boundary condition.

The following procedure is applied to find dominant splitting symmetries. During a time integration, splitting indicators $\mathcal{P}_l(\bar{t}) = \left| \int e^{i l \phi} |\bar{\Psi}(\bar{\mathbf{r}}, \bar{t})|^2 \right| {\rm d}^3 \bar{r}$ are computed at each time instance $\bar{t}$. The number $l_{\rm dom} \in \mathbb{N}_+$, for which $\mathcal{P}_{l_{\rm dom}}(\bar{t}) \ge \mathcal{P}_l(\bar{t})$, $\forall l \in \mathbb{N}_+$, indicates the dominant splitting symmetry. Vortex dynamics is divided into three categories: If $\mathcal{P}_{l_{\rm dom}}$ exceeds 0.1 before the time reaches 200, we classify the case as vortex splitting with $l_{\rm dom}$-fold symmetry (see Fig.~\ref{fig_typical}). Otherwise, if $\mathcal{S}(\bar{t}) < 0.1$ for the entire integration interval $0 \le \bar{t} \le 200$, we detect a relatively stable vortex and label this case as {\it no split}. Otherwise, we observe an unstable vortex without any obvious dominant splitting symmetry; this case is called {\it unclear}. 

\begin{figure}[h!b]
\includegraphics[width=0.96\columnwidth]{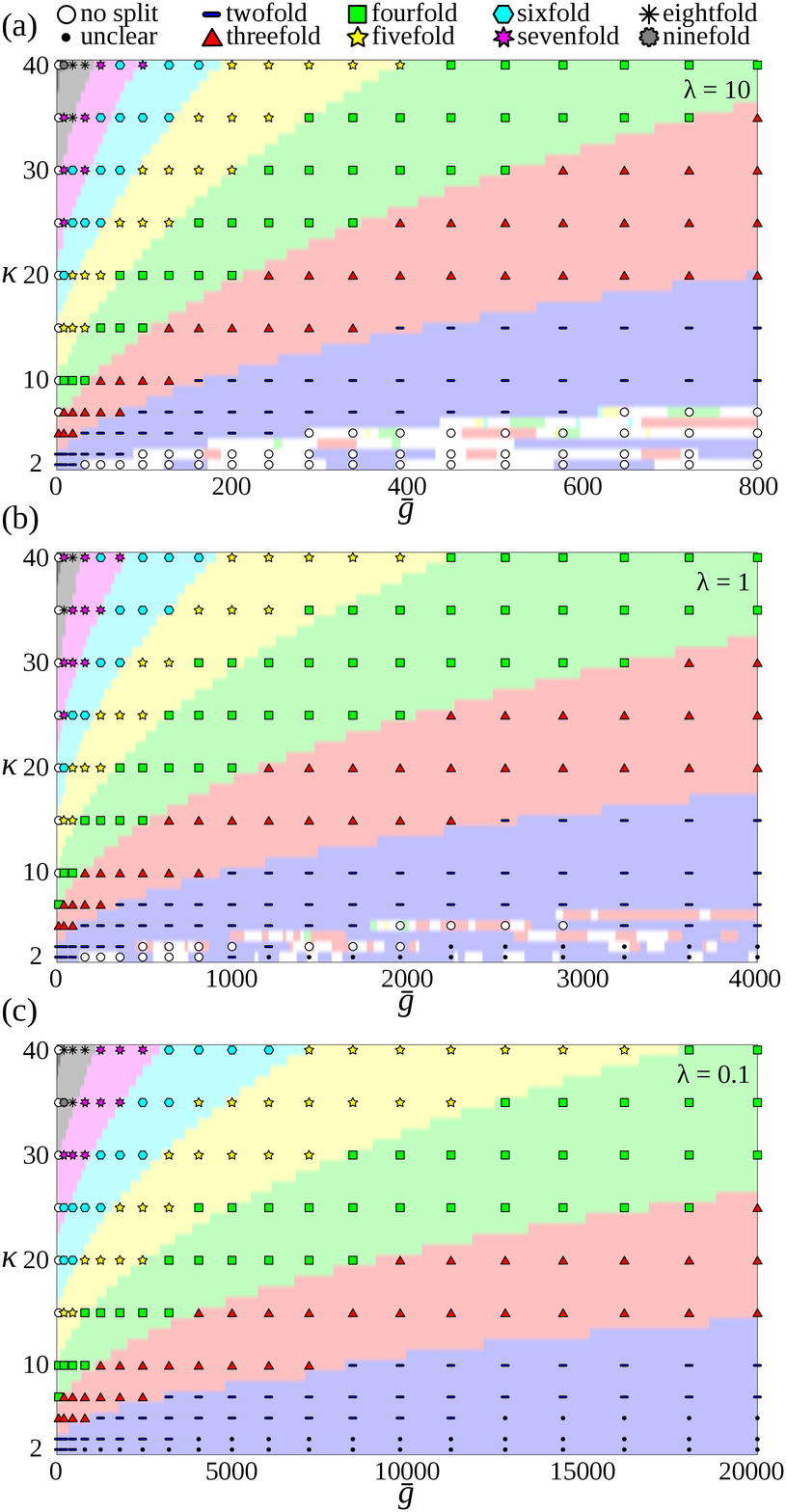}
\caption{Observed splitting symmetries in (a) the oblate ($\lambda = 10$), (b) spherical ($\lambda = 1$), and (c) prolate ($\lambda = 0.1$) condensates. The symbol indicates the result of the time integration, while the background color corresponds to the prediction of the Bogoliubov equation, namely, the value of $|l|$ for which $\max_q \mathrm{Im}(\omega_{q,l})$ is largest.}
\label{fig_splitting_diagrams}
\end{figure}

Three representative trapping ratios $\lambda$ are employed to simulate \emph{oblate} ($\lambda = 10$), \emph{spherical} ($\lambda = 1$), and \emph{prolate} ($\lambda = 0.1$) condensates. In addition, we vary the effective interaction strength $\bar{g}$ and the winding number $\kappa$ to obtain a comprehensive understanding of the splitting process. The observations from the time integrator are not entirely unique, since the results depend slightly on the seed of the random number generator. To reduce variation, we simulate each splitting process twice with different seeds and choose the splitting symmetry that is closer to the prediction of the Bogoliubov stability analysis. The splitting symmetry predicted by the Bogoliubov equation is defined as the one corresponding to the value of $|l|$ for which $\max_q \mathrm{Im}(\omega_{q,l})$ is largest. Visual inspection of Fig.~\ref{fig_splitting_diagrams} shows that the results of the time integration mostly coincide with the predictions of the Bogoliubov equation.

The characteristics of the splitting symmetries as functions of $\bar{g}$ and $\kappa$ are similar for different aspect ratios. With lower aspect ratios, a given splitting symmetry is found at higher interaction strength, which is explained by the increased size of the condensate. The most significant difference is that the unclear splitting symmetries appear only in prolate and spherical condensates. This phenomenon will be studied in more detail in the next section.

\subsection{Intertwining of vortices}

\begin{figure}[htb]
\includegraphics[width=\columnwidth]{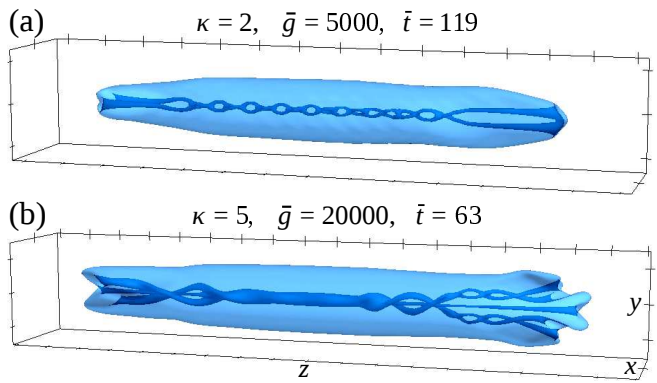}
\caption{Transparent isosurface of the particle density $|\bar{\Psi}(\bar{t})|^2$ demonstrates the intertwining of two- and five-quantum vortices in a prolate condensate ($\lambda = 0.1$).}
\label{fig_intertwine}
\end{figure}

In prolate condensates, we observe vortices to intertwine as they split, as illustrated in Fig.~\ref{fig_intertwine}. Similar intertwining processes of doubly quantized ($\kappa = 2$) vortices have already been discovered in Refs.~\cite{Mot2003.PRA68.023611,Huh2006.PRL97.110406,Mat2006.PRL97.180409}. Our study demonstrates that intertwining also occurs for large winding numbers. The branched intertwining of a five-quantum vortex ($\kappa = 5$) is illustrated in Fig.~\ref{fig_intertwine}(b).

The intertwining of vortices does not occur in the oblate condensates with the aspect ratio $\lambda = 10$, but the phenomenon seems to become observable when $\lambda$ is close to 1. To investigate this further, we consider the dynamics of three-quantum ($\kappa = 3$) vortices for different aspect ratios. To equalize the local peak interaction strengths, the effective interaction strength $\bar{g}$ is chosen to be inversely proportional to the aspect ratio as $\bar{g} = 1000 / \lambda$. 

\begin{figure}[htb]
\includegraphics[width=\columnwidth]{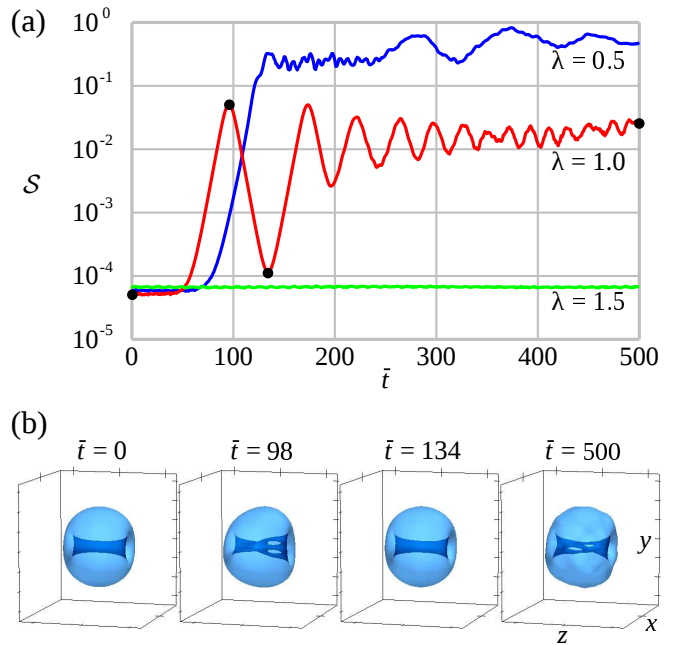}
\caption{Effect of the aspect ratio $\lambda$ on the stability of a three-quantum vortex with the interaction strength set to $\bar{g} = 1000 / \lambda$. (a)~The deviation $\mathcal{S}(\bar{t})$ from the stationary state as a function of time. (b)~Particle density isosurfaces visualizing the split-and-revival effect observed for $\lambda = 1.0$.}
\label{fig_lambda_variation}
\end{figure}

The simulations indicate that the vortices in the oblate condensates of $\lambda \ge 1.5$ are stable. In the prolate condensates with $\lambda \le 0.5$, the vortices seem to be unstable and exhibit intertwining. In between the oblate and the prolate, no prevalent behavior of the vortices is detected. Nevertheless, in a condensate with $\lambda = 1.0$, we discover a cyclic splitting process, where the vortex begins to split but then returns nearly to its initial state. This split-and-revival effect is illustrated in Fig.~\ref{fig_lambda_variation}.

\subsection{\label{sec:results4}Computational performance} 

\begin{figure}[h]
\includegraphics[width=\columnwidth]{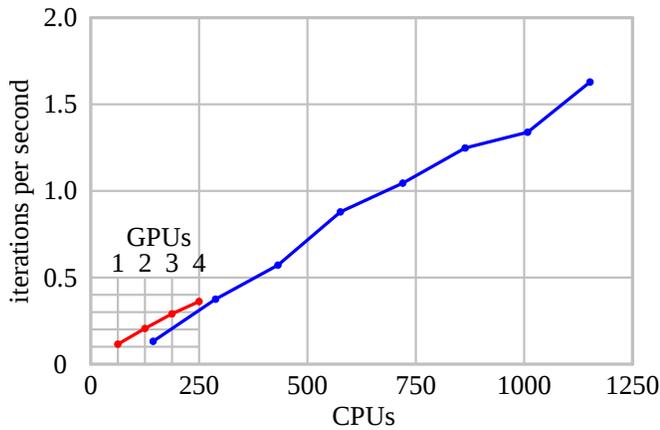}
\caption{Performance of the CPU implementation (blue) and the GPU implementation (red) as a function of the computing resources. One iteration corresponds to the integration over one unit of time.}
\label{fig_cpuAjat}
\end{figure}

The time integrations of this paper were executed on central processing units (CPUs), but we have also implemented the solver with graphics processing units (GPUs). The performances of the two implementations are studied here by measuring the simulation times in the case $\lambda = 0.1$, $\bar{g} = 5000$, and $\kappa = 20$. We use up to 96 12-core Intel (Xeon) Haswell (E5-2690v3, 64bits) CPUs and up to four NVIDIA Tesla P100 GPUs. The results in Fig.~\ref{fig_cpuAjat} indicate that the performance of the GPU implementation on one GPU corresponds to the performance of the CPU implementation executed on at least 60~CPU cores.

\section{Conclusion}\label{sec:discussion}

In summary, we have studied the splitting dynamics of giant vortices in dilute BECs with a particular focus on the time integration of the three-dimensional GPE. We showed that a significant reduction of the numerical error is achieved when a tetrahedral spatial tiling is utilized instead of the routine Cartesian grid. Importantly, the careful choice of the numerical method provides us with the physically correct splitting symmetry.

Comprehensive maps of vortex splitting symmetries were presented for oblate, spherical, and prolate BECs. The solutions of the time integrations were found to agree with the linear stability analysis based on the Bogoliubov equation.

The splitting-induced intertwining of vortices in prolate condensates is demonstrated. The aspect ratios for which the intertwining becomes observable are also studied. A split-and-revival phenomenon, where the vortex almost returns to its initial state after splitting temporarily, was observed in the crossover from a dynamically stable vortex into an unstable one as a function of the aspect ratio.

The performance study presented in Sec.~\ref{sec:results4} indicates nearly optimal scalability of the CPU implementation and promising performance for the GPU implementation. In the future, we will study how the GPU performance scales with a larger number of GPUs. This will allow us to accomplish even more challenging tasks than is currently possible with CPUs. These tasks may include solving the dynamics of a lattice of monopole--antimonopole pairs~\cite{2014ray,2015ray}.

\begin{acknowledgments}
We have received funding from the European Research Council under Consolidator Grant No. 681311 (QUESS), the Technology Industries of Finland Centennial Foundation, and the Academy of Finland through its Centres of Excellence Program (project Nos. 312300, 251748, and 284621) and Grant No. 308632. The computing resources were provided by the CSC - IT Center for Science, which is owned by the Finnish Ministry of Education and Culture.
\end{acknowledgments}

%
%\bibliography{viitteet,additional_refs}% Produces the bibliography via BibTeX.
\end{document}